\documentclass[preprint2]{aastex}

\begin{document}


\title{AAT Imaging and Microslit Spectroscopy in the Southern Hubble Deep Field }


\author{Karl Glazebrook{\altaffilmark{1,2}}, 
Aprajita Verma\altaffilmark{3,4},
Brian Boyle\altaffilmark{5,2}, 
Sebastian Oliver\altaffilmark{3,6}, 
Robert G. Mann\altaffilmark{3,7},
Davienne Monbleau\altaffilmark{2}
}


\altaffiltext{1}{Department of Physics and Astronomy, Johns Hopkins University, Baltimore, MD 21218}
\altaffiltext{2}{Anglo-Australian Observatory, P.O.Box 296, Epping NSW, Australia}
\altaffiltext{3}{Astrophysics Group, Imperial College London, Blackett Laboratory, Prince
Consort Road, London, SW7 2AZ, UK}
\altaffiltext{4}{Max-Planck-Institut f\"ur extraterrestrische Physik, Giessenbachstra{\ss}e, 85748 Garching, Germany}
\altaffiltext{5}{Australia Telescope National Facility, PO Box 76, Epping NSW 1710, Australia}
\altaffiltext{6}{Astronomy Centre, University of Sussex, Falmer, Brighton BN1 9QJ, UK}
\altaffiltext{7}{Institute for Astronomy, University of Edinburgh, Royal Observatory, Blackford Hill, Edinburgh EH9 3HJ, UK}


\begin{abstract}
  
  We present a deep photometric ($B$- and $R$-band) catalog and an
  associated spectroscopic redshift survey conducted in the vicinity
  of the Hubble Deep Field South. The spectroscopy yields 53
  extragalactic redshifts in the range $0<z<1.4$ substantially
  increasing the body of spectroscopic work in this field to over 200
  objects.  The targets are selected from deep AAT prime focus images
  complete to $R<24$ and spectroscopy is 50\% complete at $R=23$.
  There is now strong evidence for a rich cluster at $z\simeq 0.58$
  flanking the WFPC2 field which is consistent with a known absorber
  of the bright QSO in this field. We find that photometric redshifts
  of $z<1$ galaxies in this field based on HST data are accurate to
  $\sigma_z/(1+z)=0.03$ (albeit with small number statistics).  The
  observations were carried out as a community service for Hubble Deep
  Field science, to demonstrate the first use of the `nod \& shuffle'
  technique with a classical multi-object spectrograph and to test the
  use of `microslits' for ultra-high multiplex observations along with
  a new VPH grism and deep-depletion CCD.  The reduction of this new
  type of data is also described.

\end{abstract}



\section{Introduction}

\def\todo#1{{[\bf TODO: #1]}}  \def\REF{\todo{REF}}

The Hubble Deep Field South (HDF-S) is one of the deepest
imaging fields in the sky. It was observed in 1998 (Williams et al.
2000) by the Hubble Space Telescope (HST) as a counterpart to the
northern Hubble Deep Field (HDF-N). However in contrast to the
northern field it was selected to contain a bright QSO J2233-606 at
$z=2.24$ in order to facilitate studies of the connection between
foreground galaxies and absorbing systems in the QSO spectrum. 

The determination of precise redshifts for extra-galactic sources has
been important since the time of Hubble (1929). While much spectroscopy
has been done in the other Hubble Deep Field, HDF-S has lagged behind.
As a community service and in order to test new instrumentation
techniques, in particular `nod \& shuffle' (Glazebrook \&
Bland-Hawthorn 2001; GB01) with very small `microslits', we carried
out a spectroscopic campaign of galaxies in, and in the vicinity of,
the HDF-S in order to obtain redshifts.  The plan of this paper is as follows: Section 2
  details the prime focus pre-imaging, the procedures
  used to construct the photometric catalog 
  and the selection of the spectroscopic
  targets. In Section 3 we describe our novel spectroscopic
observations and the special data reduction procedures used. In
Section 4 we present our spectroscopic catalog and its basic
parameters and compare with other work in this field and discuss the
possible galaxy cluster at $z\simeq 0.58$. 

The finding charts, images, spectra, photometric and spectroscopic catalogs presented
in this paper are all available from the Anglo-Australian Observatory
web site at:

{\tt http://www.aao.gov.au/hdfs/Redshifts}

\section{Imaging Observations \& Spectroscopic Selection}

\begin{figure*}[!h]
\begin{centering}
 \includegraphics[width=15cm,angle=90]{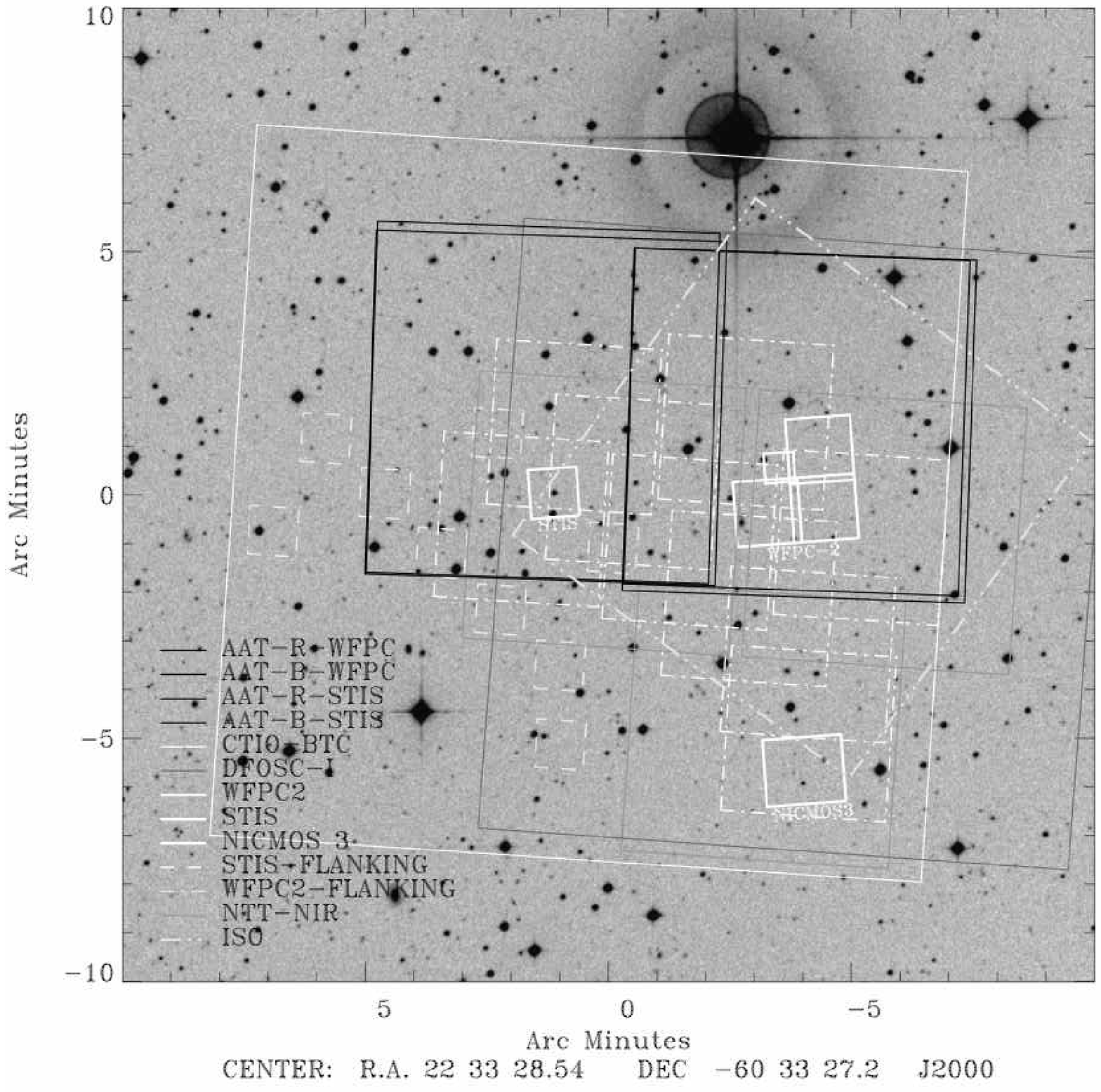}
\caption{Figure depicting follow-up imaging surveys of the HDFS. The extent of the AAT images is shown in black. The bright star HD 213479 which causes stray
light in the AAT fields can also be seen. 
\label{star_fig}
}
\end{centering}
\end{figure*}

We obtained pre-imaging data\footnote{We briefly describe the imaging
  here and further details are given on the project website.} (prior
to the HST campaign) from which our sample of targets for
spectroscopic follow-up were selected.  
Images in $B$ and $R$ were taken with the Anglo Australian
Telescope Prime Focus CCD camera (0.391 \arcsec$/$pixel) in May 1999 at two pointings
containing the WFPC (`AAT-WF') and STIS (`AAT-ST') fields respectively.
When stacked and
mosaicked, they cover a contiguous area of $12.5'\times7'$ centered on 22$^{\rm h}$ 33$^{\rm
  m}$18.68$^{\rm s}$ $-$60$^\circ$31$\arcmin$45.8$\arcsec$ (J2000)) which includes the HST deep fields
  and their flanking fields. All images were taken in photometric conditions and 0.8\arcsec\ seeing,
 except for the $B$-band AAT-ST data (2.2\arcsec\ through thick clouds).
 A finding chart for our pointing is given in Figure~\ref{star_fig}.

The images were de-biased, flat-fielded and mosaiced following
standard CCD procedures. All images were
registered to the USNO-A2.0 reference frame using x- and y-shifts and
rotation. The internal astrometric offset between sources in
overlapping frames is measured to be $\approx$ 0.12\arcsec.
An all-sky photometric solution was
determined from observations of 7 Landolt (1992) fields using 29
stars.  $B$-band images taken in non-photometric
conditions and were calibrated using overlap areas with photometric
images (accurate to $<0.04$ mags residual scatter). A 
source catalogue was extracted using Sextractor version 2.2.1 (Bertin and
Arnouts, 1996).  Sextractor was run in the standard manner using a low
detection threshold of 1$\sigma$ and a minimum of 3 connected pixels
to define the extraction. We used a local determination from the sky
to accurately handle data with variable sky levels. This is
particularly important since the images suffer from strong stray light
pollution from an off-image star to the north of the field (HD 213479
at 22$^{\rm h}$ 33$^{\rm m}$10.97$^{\rm s}$
$-$60$^{\circ}$26$\arcmin$00.4$\arcsec$, $B$=8.1). In order to
maximise the reliability of the final catalogue without compromising
the faint source detection, we flagged sources which lie in areas most
strongly affected by straylight or in vignetted regions as `masked' to
minimise the number of spurious extractions. Only sources lying within
unmasked regions are considered as targets and for further
analysis.  The magnitudes presented in the final catalogue are
SExtractor {\tt MAG\_AUTO} magnitudes determined using an automated
adapative aperture technique. The final photometric catalog is given in Table~\ref{tab:phot}
(electronic edition only).

  We extracted an extragalactic catalogue using star-galaxy separation
  and analysed the resulting extragalactic number counts. We chose to use the
  object's magnitude and a measure of surface brightness
  (magnitude/FWHM$^{2}$) as our star-galaxy separator as it provides a
  cleaner cut at faint magnitudes than SExtractor's
  classifier or the commonly used core:total indicator.  
  Our number-magnitude analysis showed the counts of galaxies to be
  still increasing as a power-law to $R$=24, consistent with close to
  100\% completeness at this depth. The formal magnitude limit of the images, 
  measured from the noise in random 0.8\arcsec\ diameter apertures
  placed on blank regions is $R$=25.30, 25.16 (WF, ST) for a 3$\sigma$
  detection. This is also
  consistent with the images being highly complete at $R=24$.
  The $B$-data reaches 25.18 (WF) and 23.97 (ST, 2\arcsec\
  aperture --- due to the poorer seeing). At $R=24$ galaxies overwhelmingly
  dominate stars  and so we chose to select
  spectroscopic targets from a purely $R$ magnitude limited catalog to
  this depth.  $R=24$ was somewhat deeper than we expected to reach with
  spectroscopy even under the best conditions, however we expected to
  go deeper for objects with strong emission lines and we also desired
  a high surface density to test the `microslit' spectroscopy
  described below so our final strategy was to observe a lot of
  objects and tolerate a lower completeness. The final list of objects
  that made it on to the slit-mask was based solely on geometrical
  considerations.
  
  Finally, we note that the spectroscopic selection was done on a
  preliminary version of the final astrometric and photometric table
  presented here. We find that the positions and magnitudes reproduce
  well between the preliminary catalog and the final version ($<0.4$
  arcsec and $<0.03$ mag systematic differences).

\section{Spectroscopic Observations}

The spectroscopic observations were made with the Anglo-Australian
Telescopes Low Dispersion Survey Spectrograph (Wynne \& Worswick
1988).  This instrument (LDSS) was designed for multi-slit imaging
spectroscopy and offers a wide circular (12.3 arcmin diameter) field
of view. Data was taken in October 1998 during a
commissioning run to test the following new features:

\begin{enumerate}
\item A new deep-depletion CCD detector from MIT Lincoln Labs
 (see Burke et al. 2004 for details on these devices) had just
  been installed offering improved red sensitivity. The deep depletion
  of high-resitivity p-type Silicon  causes pixels to be up to 40\micron\ deep, this allows
  more depth for a photon to be detected and improves the quantum
  efficiency $>8000$\AA\ where Silicon is becoming increasingly transparent.
  It also reduces the effect of fringing. We obtained a 
   $2048\times 4096$ pixel device (15 \micron\ pixels giving 0.395
  arcsec $/$ pixel). The red
  quantum efficiency peaked at 87\% at 7000\AA\ and maintained 16\% at
  10,000\AA. (Tinney \& Barton 1998).
\item A new red-optimized grism based on a Volume Phase Holographic
  (VPH) grating from Kaiser Optical Systems was tested. VPH gratings offer
  improved throughput (Barden et al. 1998). This was the first test of
  a low-resolution VPH grating designed for redshift measurements in a
  multi-object system and proved successful. The peak efficiency was
  measured to be 82\% at 6700\AA\ (Glazebrook 1998). The grating
  delivered a dispersion of 2.6\AA\ per pixel with the MIT/LL CCD and
  a one arcsec slit projected to $\sim 3$ pixels giving 8\AA\ spectral
  resolution.
\item The use of `nod \& shuffle' (GB01) for accurate sky-subtraction
  for the first time in a multi-object configuration.
\item The use of very small slits (`microslits'), enabled by nod \&
  shuffle, to allow large multiplex, as described below.
\end{enumerate}

\subsection{Nod \& Shuffle}

The nod \& shuffle technique for highly accurate
sky-subtraction is extensively described in GB01. In a
nutshell the technique uses an unilluminated part of the CCD as a
storage area for an image of the sky. Observing consists of taking an
image of the objects through the slits, clocking the CCD voltage pattern so
this image is `shuffled' in to the storage, nodding a few arcsecs, taking
an image of the sky through the slits, and repeating. The result is
quasi-simultaneous images of the object and sky adjacent on the CCD.
This is done rapidly to track sky variations, e.g. for HDF-S we took 30 sec images at
object and sky positions (plus a 2 sec deadtime  to allow
for telescope settling). The sequence is repeated 30$\times$ 
giving an 1800 sec (900 sec on object)
exposure. Readout noise only becomes important during the final
clock-out, charge shuffling is essentially noiseless as long as the
CCD has good charge-transfer efficiency (CTE). Sky-subtraction then consists
of windowing the sky region of the image and then subtracting it from
the object region which is offset by an exactly known number of pixels.
With the setup employed here the systematic sky residuals do not exceed 0.04\%
(measured in GB01).

\begin{figure*}[!h]
\epsscale{1.5}
\plotone{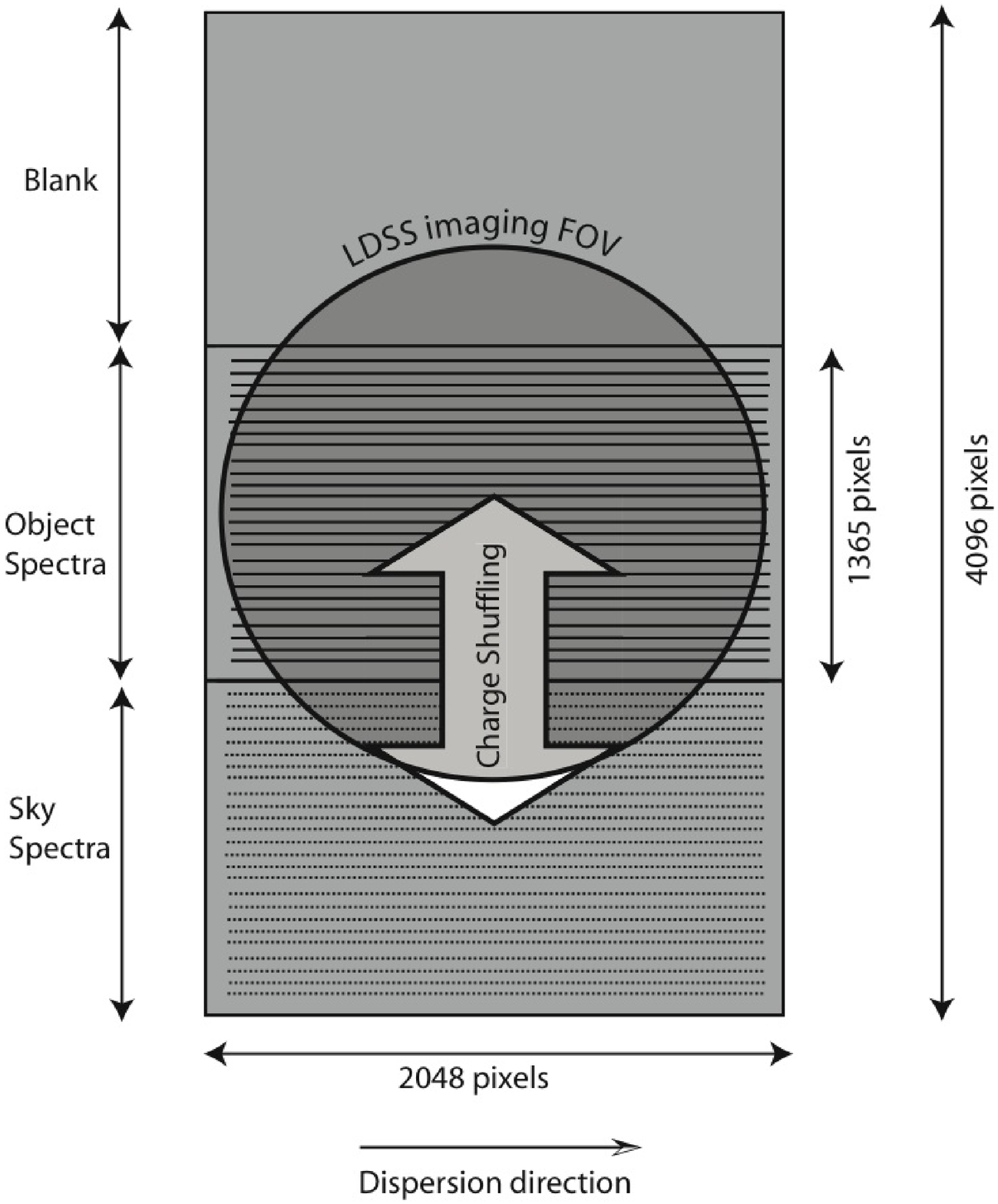}
\caption{Illustration of the particular LDSS nod \& shuffle layout we used. The $2048\times 4096$ CCD detector (grey rectangle) is physically larger than the
  circular image (12.3 arcmin diameter) delivered by the LDSS camera.
  For nod \& shuffle only the middle third of the chip (1365 pixels
  high) is illuminated by the mask, the lower third is used to store
  the sky image and the upper third is blank (it hold the object image
  when the shuffling charge up and down by 1365 pixels). Thus only 9.0
  arcmin vertically is available for slits which is almost the full
  field.}
\label{fig-ns}
\end{figure*}

Many schemes are possible for laying out positions of slits and
storage areas on the detector. Whether one can shuffle a long
distance or a short distance depends on the CTE and the number
of charge traps on the detector (each trap causes a trail of charge
when shuffled). 
For example Abraham et. al. (2004)
in their later Gemini Multi-Object Spectrograph (GMOS)  observations 
used 2 arcsec long rectangular slitlets, and `microshuffled' the charge only
one slit length. The storage was immediately below
the detector so effectively 50\% of the GMOS FOV was available to
slits. For these small shuffles CTE is not critical, also the GMOS CCDs
had hundreds of charge traps so small shuffles was desirable.

For our observations the MITLL3 CCD only had 12 traps and the 
CTE was 99.9999\%
which allowed us to make large `macroshuffles' across the whole
device. 
We chose to  follow the scheme shown in Figure~\ref{fig-ns} which is
motivated by the fact that the MITLL3 device
is physically much larger than the LDSS focal plane. One third
of the device (1365 pixels = 9.0 arcmin) is used for imaging, the rest
for storage. 75\% of the LDSS FOV is available for slits, which is advantageous,
but we have to shuffle 1365 pixels. This is not a problem with the high CTE, tests
in the lab showed significant image degradation did not happen 
until several hundred macroshuffle operations had been performed. Finally, we note
that the FOV advantage in macroshuffling 
is only gained when the FOV greatly underfills the CCD
and would not be applicable, for example, to GMOS.  

\subsection{Microslits \& Mask design}

The unconventional approach taken with the spectroscopic masks was
that instead of using rectangular slitlets we used small circular
holes in the mask at the location of target galaxies. The logic was
that the usual reason to employ an extended rectangular slit several
arcsecs long is to sample neighboring sky for sky-subtraction. With
the nod \& shuffle technique this need is obviated. Thus one only
needs apertures large enough to receive the light from typical targets
in the expected seeing. Normal LDSS slitlets are of order 5--10 arcsec in
length, thus with 1 arcsec `microslits' we could obtain a
5--10$\times$ larger multiplex.

Ideally we would have used square holes so that the spectral PSF would
be uniform across the target but due to hardware limitations of our particular
mask cutter it turned out to be much easier to simply drill circular
holes. The PSF variation is not important for one dimensional spectra
summed across the slit and does not affect nod \& shuffle sky
subtraction as objects and sky are observed through exactly the {\em
  same} slits (this is confirmed by the essentially Poisson-limited sky
  subtraction in the final spectra).  Note we planned to do the observations in the best AAT
seeing ($\la1$ arcsec) and so we cut one arcsec holes (150 \micron\ 
diameter). The masks were available immediately before the run but
there was no opportunity to measure them until after the run. Using a
microscope we later measured the diameters to be on average 0.7
arcsec, this would have resulted unfortunately in light loss given the
actual seeing. This non-optimal size lessened our final spectroscopic
depth.

Slit allocation was done using a custom algorithm which simply
maximizes the number of non-overlapping target spectra for a given
position angle. We chose $90^{\circ}$ (i.e. along the STIS--WFPC2
axis) in order to maximize the number of targets allocated within
the Hubble camera fields. We found the typical overall number of slits
allocated by our algorithm was not sensitive to this choice.
The position angle is not optimal for atmospheric dispersion,
however since we were doing red spectroscopy this would not have been a
significant source of light loss.  Using the red VPH grating the 2048
pixels in the dispersion (CCD row) direction gives 5300\AA\ wavelength
coverage. The typical wavelength range observed for a slit on the mask
center was 5000--10000\AA. Slits were allowed to be as far
as 1.5 arcmin off-axis resulting in wavelength coverage shifts of up
to 590\AA. Thus it was only possible to accommodate one object at each
CCD row (i.e. one tier). Slit allocation allowed a minimum one pixel gap
between objects to cleanly delineate spectra, thus the maximum
possible number of holes was 348. Running the algorithm on the $R<24$
HDF-S catalog for a mask center of 22$^h$ 33$^m$ 11$^s$ $-60^\circ$
33$'$ 16$''$ gave an allocation of 225 objects (after taking out some
space to allow for larger holes for alignment stars) in the $3$ arcmin
$\times$ 9 arcmin field. This compares with a normal LDSS mask setup
without nod \& shuffle where one normally only gets spectra of 20--30
objects.

The drawback of this strategy is that the objects are out of the slits
for half the time when the telescope is in the sky position. An
alternate nod \& shuffle approach would be to use slightly larger slitlets so the
object is still in the slit in both nod positions (e.g. Abraham et al.
2004). However this means in
general one can only observe half as many objects because the slits
are twice the spatial size, for a {\em high sky density survey\/} there is no net
difference in the number of objects observed to a given depth in a given
amount of telescope time.

\subsection{Observations}

The observations were carried out over the nights of 13--16$^{\rm th}$
October, 1998 at the AAT. Conditions were photometric and the seeing
was 1--1.5 arcsec.  The red VPH grating was used with a GG475 blocking filter,
so the spectra are potentially contaminated by second order for $>$ 9000\AA, though
this effect was not in practice seen in any spectra. The
filter was also not anti-reflection coated causing some ghosts in the spectral images.

The total exposure time obtained on the HDF-S field was 12 hours (half
on sky). The main calibrations taken were arc spectra for wavelength
calibration and white-light
dispersed flat fields (through the mask). The flatfields served the
dual purpose of removing pixel sensitivity variations and allow the
spectra location on the detector to be mapped empirically.

\subsection{Data Reduction}

\begin{figure*}[!h]
\epsscale{1.8}
\plotone{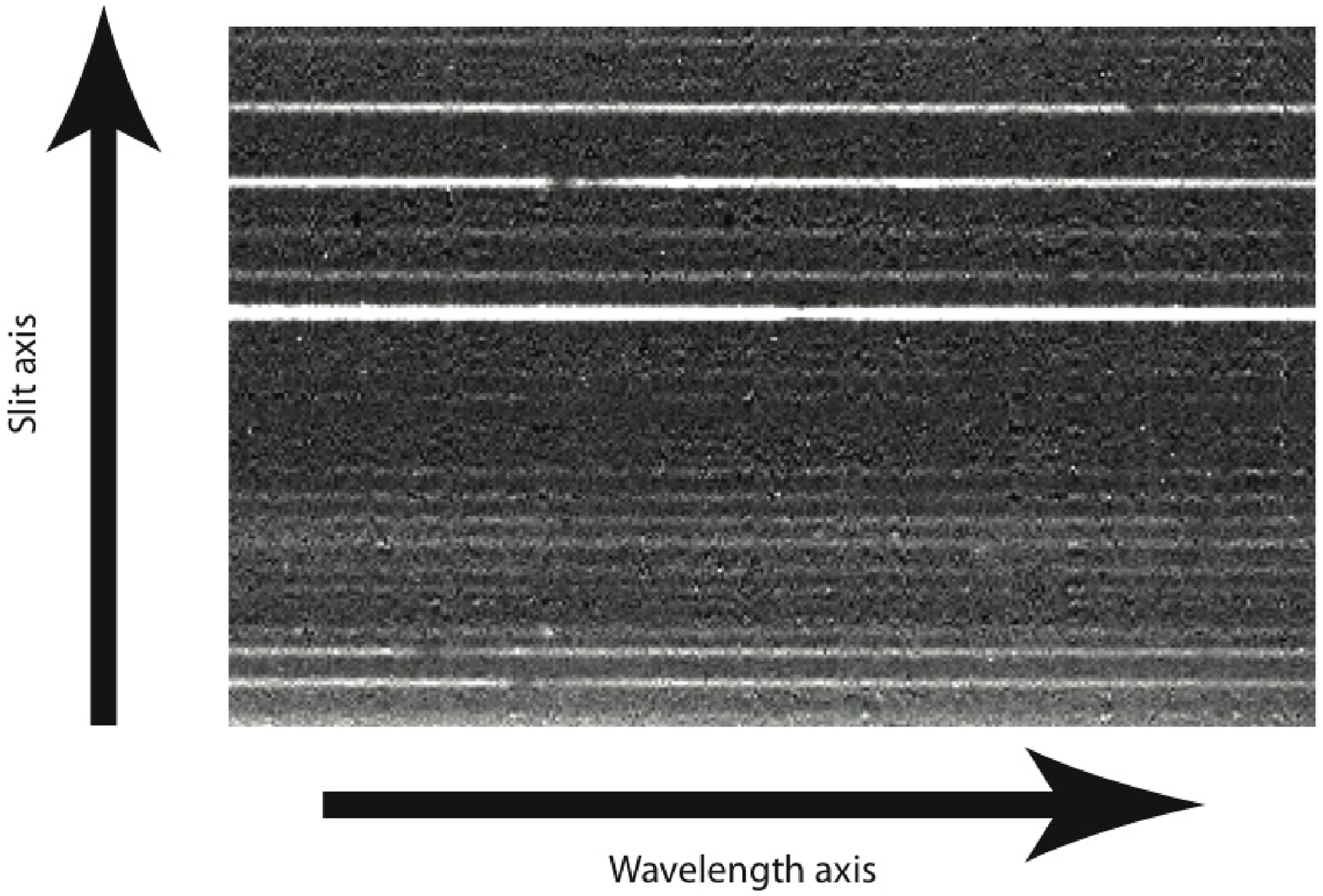}
\caption{Example microslit data. This shows a zoom on a small portion of a 2D data frame ($\sim 30$ microslits) where the only processing
applied at this step is the subtraction of the shuffled sky. Each circular microslit produces a spectral trace. Several bright objects with absorption
lines and faint objects with emission lines can be seen. The data shares features both with fiber data and classical multislit data. The apertures are unresolved spatially, however
unlike fiber data the spacing between apertures is uneven (it reflects the spacing between objects) and the wavelength zeropoints vary too
(reflecting the spatial position of the objects on the other axis).}
\label{fig-datared}
\end{figure*}

The format of the data shares similarities both with classical
multislit spectra and with fiber spectra. Each small circular aperture
produces a spectral trace (`tramline') on the CCD image. A portion of
this data is shown in Figure~\ref{fig-datared}. We reduced the data using a mixture of
{\it IRAF} (Tody 1986) tasks and custom {\it Perl Data Language\/} (Glazebrook \& Economou 1997)
scripts. First the individual images were sky-subtracted, using the shuffle
  offset to map the sky pixels to the object pixels. The images (+ arcs and flats) were then corrected for small
  alignment shifts throughout the night and registered and stacked with a cosmic ray filter.
The flat-field was used to map the spectra tramline locations and also the PSF width as a function of wavelength for optimal extraction
of 1D object, sky (for reference), arc and flat-field spectra.  The flat-field spectra were then normalized with a polynomial fit in the spectral
  direction and divided in to the object spectra to correct  them for high frequency pixel/wavelength
  sensitivity variations. An overall wavelength solution was derived
  from the arc and sky spectra ($1$\AA\  RMS).

Care was taken to propagate variance arrays (with initial errors calculated from the 
CCD readnoise and gain using photon statistics) through every step of the data
reduction so errors could be assigned to the final spectra. 

One important point about the data reduction is that the shuffled sky must be subtracted {\em before}
flat-fielding. This is a key advantage of nod \& shuffle: accurate sky-subtraction does
not require flat fielding. In the shuffled image the relevant pixel response is that of the original 
pixel and not the storage pixel. So the correct procedure is to subtract the sky first.
The pixel response is the same and the sky subtracts correctly. Then applying
the flat field will correct the flat field error in the object correctly. If the flat field is
imperfectly measured this only affects the object spectrum. Reversing the procedure,
i.e. flat fielding before sky subtraction will actually result in an erroneous flat field
correction. 

\section{Spectra and Redshift catalog}

Redshifts were determined from the final set of 225 target spectra by
careful visual inspection and are given in Table~\ref{tab-spec} along with key spectral features used
for identification and a subjective quality system --- $Q=4$ denotes a dead
certain redshift ($>99$\% confidence), $Q=3$ is `quite certain' ($\sim
80$\% confidence), $Q=2$ is a `possible identication' ($\sim 50$\%
confidence), $Q=1$ is a single emission line redshift assumed to be
[OII] (which is the best a priori guess given the likely redshift
range at $R<24$ and wavelength coverage) and $Q=0$ represents no identification.

\begin{figure*}[!h]
\epsscale{1.8}
\plotone{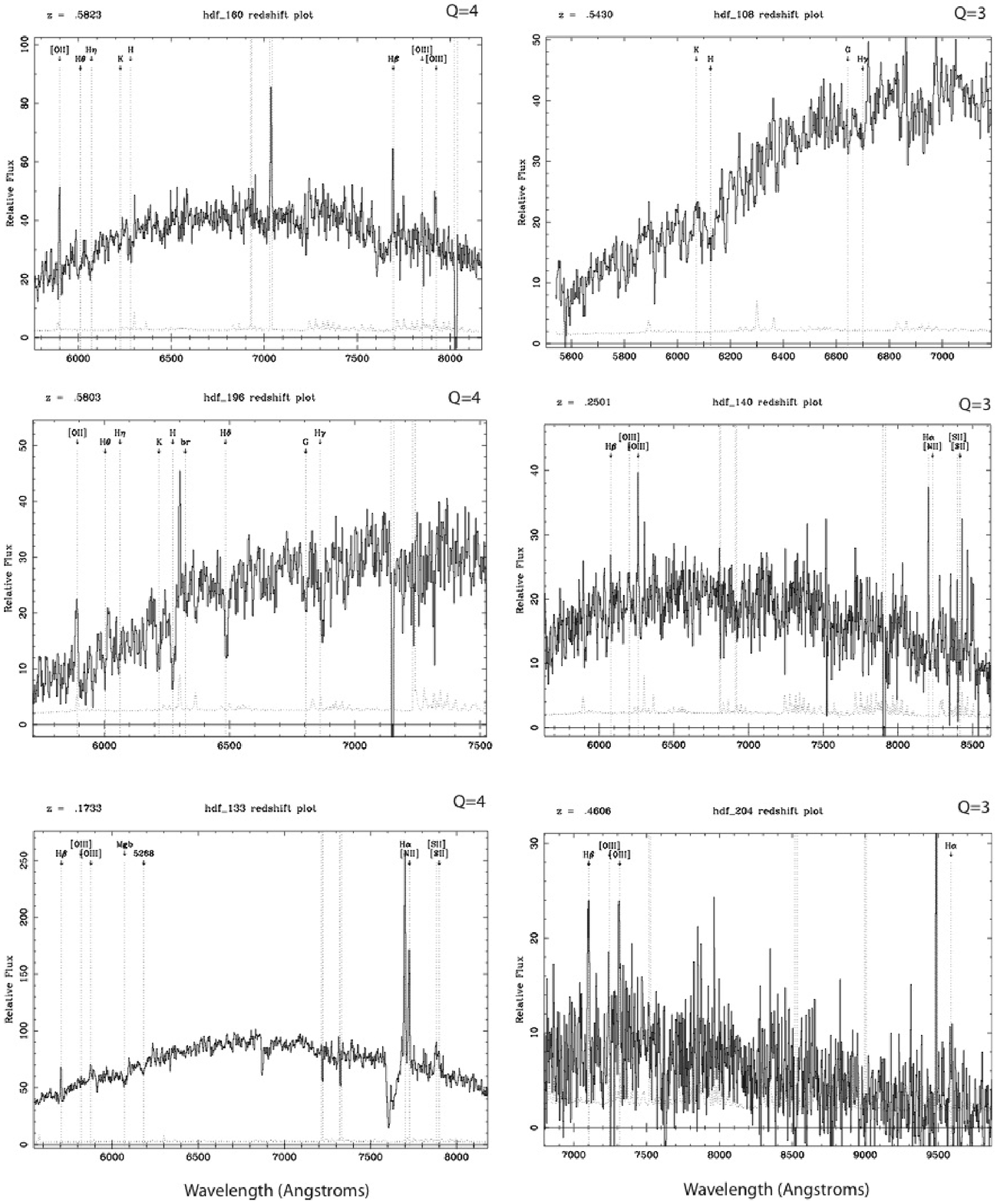}
\caption{Random spectra selected from the survey to illustrate the quality of the spectra.
Spectra with $Q=4$ are shown in the left column and $Q=3$ are shown in the right column.
Identifying spectral features are labeled. The spectra are unfluxed and are shown in counts.
The dashed line is the 1$\sigma$ noise amplitude spectrum calculated from the CCD 
properties.}
\label{fig-spec}
\end{figure*}

Out of the 225 objects there were (24, 19, 22,8, 152) objects with
$Q=(4,3,2,1,0)$ respectively. Thus 73 identifications (33\%) were made
of which 20 were galactic stars. The completeness was 79\%, 63\%,
46\% at $R<$ 21, 22, 23 respectively. Many of the faint
$R>23$ identifications were for objects with strong emission lines.
Random example spectra of $Q=4$ and $Q=3$ objects are shown in 
Figure~\ref{fig-spec} to illustrate the quality. FITS format spectra (and variance spectra)
are available on the web site. In the final list we identify five
galaxies and two stars within the deep WFPC-2 field and one galaxy
in the STIS field. 
 
 \subsection{Comparison with other HDF-S redshifts}
 
 Spectroscopy in the HDF-S vicinity has also been done by Tresse et
 al. (1999; T99), Vanzella et al (2002; V02) and Sawicki \&
 Mall\'en-Ornelas (2004; SM03).  T99 published redshifts obtained with
 the NTT for nine objects within 1 arcmin of the QSO with $I<22.2$,
 three of these are on our mask of which we identify one (\#20) as a
 star, however this is only a $Q=2$ confidence. The other two we do
 not identify.  V02 obtained 50
 redshifts using the VLT in the range $I_{AB}=20$--25 (75\% complete
 at $I_{AB}<22.5$) with an emphasis on color selected $z>2$ galaxies.
 We find three identifications in common, all agree at the $\Delta z
 \le 0.001$ level.  SM03 obtained 97 `secure' redshifts for sources
 with $I_{AB}<24$ using the VLT. Five of their objects are also identified
 by us, all are high confidence redshifts (except for one single
 emission line redshift) and all agree to within $\Delta z \le 0.001$.
 There is one object (\#204) which appears both in V02, SM03 and in
 our catalog, all at the same redshift.  If we exclude the objects in
 common with the other spectroscopic catalogs we find we have
 identified 45 new extragalactic objects in the HDF-S and flanking fields.
 Excluding all objects in common there are now 206 unique galaxies 
 with spectroscopic redshifts
 from all the HDF-S catalogs considered here.

Another interesting comparison is with catalogs based on photometric
redshifts. Here the WFPC-2 field allows accurate photometric redshifts
because of the depth and 3 colors.  In order to compare we define the
fractional error on the redshift as $x=z/(1+z)$, comparing with the
catalog of Gwyn (1999) we find that $\sigma_x = 0.03$ between the
catalogs (5 galaxies, 2 stars).  We also compare with the
`Stony-Brook' catalog\footnote{This data is publicly available from\\ {\tt
  http://www.astro.sunysb.edu/astro/hdfs/wfpc2}} which is based on the HDF-N methods of 
  Fern{\'a}ndez-Soto et al. (1999).  Here we find
 a big discrepancy in object  \#871 which Stony-Brook identifies 
 as a star and we identify as a $z=0.7$ galaxy, albeit with very
 low confidence ($Q=2$). Excluding this object we again
 find $\sigma_x = 0.03$ between the catalogs (4 galaxies, 2 stars). 
 
 \subsection{Redshift distribution and rich cluster near WFPC2 pointing}

 \begin{figure*}[!h]
\epsscale{1.8}
\plotone{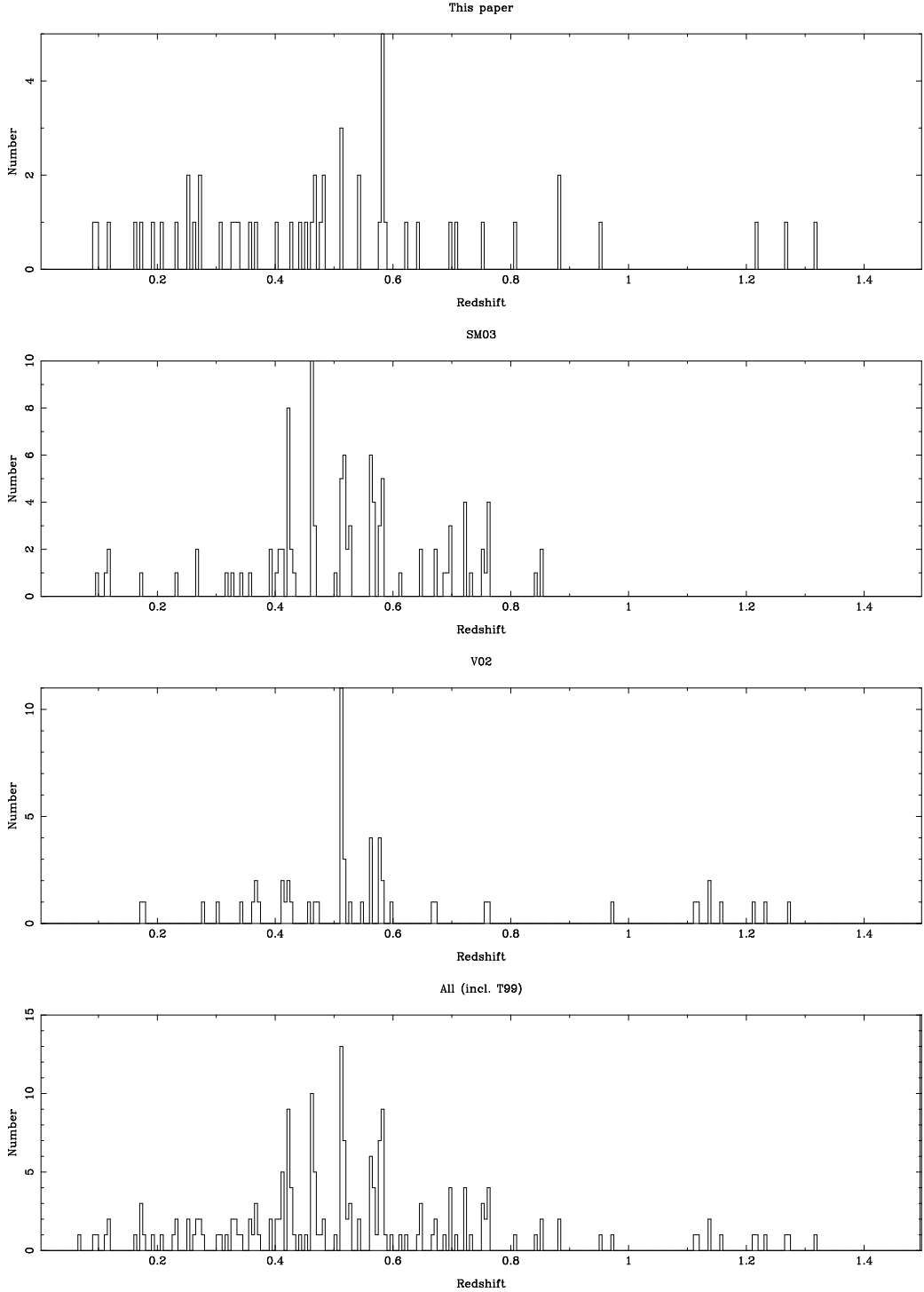}
\caption{Redshift distribution of the HDF-S galaxies in this paper compared with the V02 and SM03 samples. At the bottom the combined redshift
  distribution (also including the nine T99 objects) is shown (with all objects duplicated between catalogs removed). }
\label{fig-nz}
\end{figure*}  

\begin{figure*}[!h]
\epsscale{1.8}
\plotone{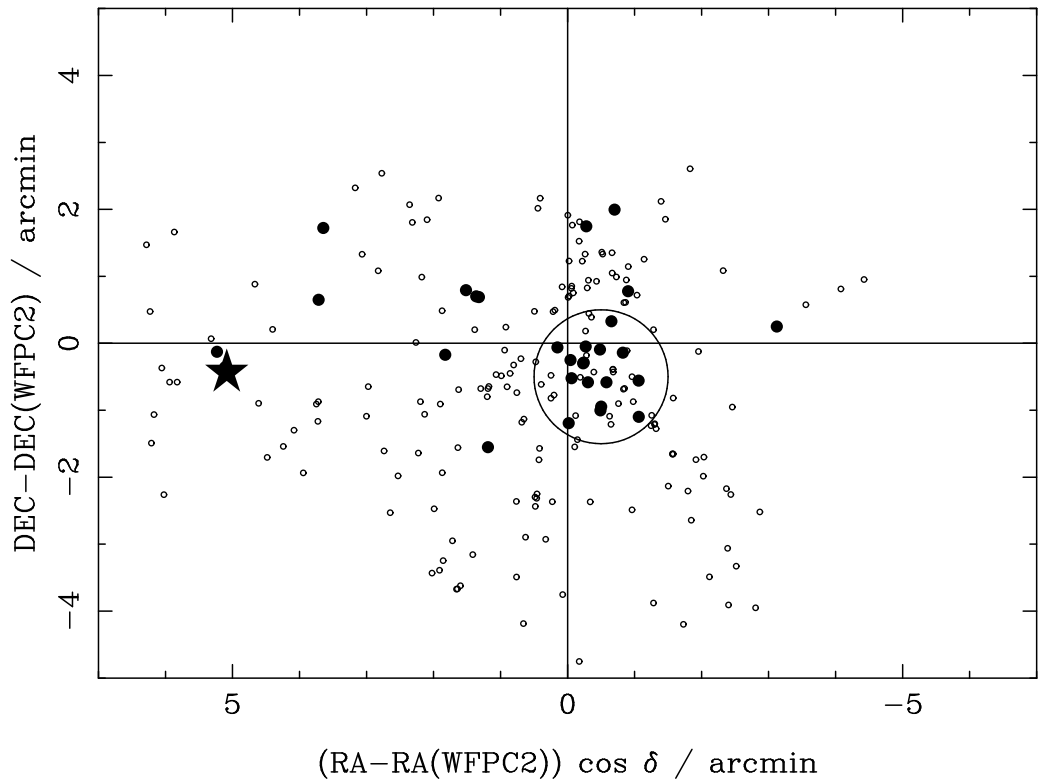}
\vbox{
 \  \\
 \  \\
 \  \\
\plotone{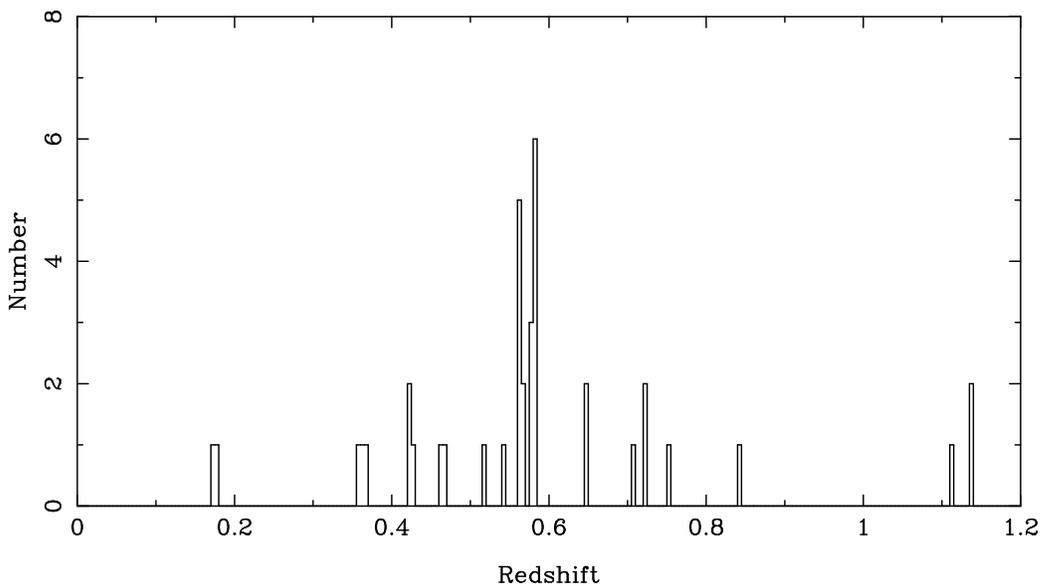}
}\caption{Top: Locations of galaxies with $0.56<z<0.60$ (solid circles) 
compared to all galaxies (open circles) with spectroscopic 
redshifts in the combined sample of this work, T99, SM03,  and V02. The circle
shpws the region used to estimate the velocity dispersion. The position of the STIS QSO is marked with a large star.
Bottom: redshift distribution of galaxies within the circle showing the strong spike at $z\simeq 0.58$. 
}
\label{fig-sky}
\end{figure*}

The resulting redshift distributions are compared in Figure~\ref{fig-nz}. V02 reported 
an over-density near the WFPC2 pointing at  $z\simeq 0.58$ 
after combining our data (based on our initial WWW report) with
theirs. It can be seen that the redshift spike is at it's  most prominent in our dataset. If we look at the
combined redshift distribution in Figure~\ref{fig-nz} it can be seen that there  are a number of spikes, as is typical for 
narrow field redshift surveys, and that the $z=0.58$ spike is not the most prominent of them.
However the STIS QSO displays a
strong absorption line system at $z=0.570$ (T99) coincident in 
redshift with a bright spiral galaxy so it is interesting to explore this further. This can be
done by looking at the distribution on the sky. Figure~\ref{fig-sky} shows a comparison 
of the locations of the galaxies at $0.56<z<0.60$ compared to all the galaxies from this work, SM03, T99 and V02 (after removing all duplicate objects).  It is clear that there is a compact cluster of 
galaxies at this redshift flanking and overlapping with the SW corner of the WFPC2 field. 
The cluster is of order one arcmin in size which is fairly typical for large clusters at these redshifts. 
Taking an approximate cluster center of $(-0.5,-0.5)$ arcmin 
wrt the WFPC2 field we count 16 galaxies within a one arcmin radius and with $0.56<z<0.60$. The redshift distribution of galaxies in this circle is
shown in the lower half of Figure~\ref{fig-sky} and it
can now be seen that there is again a single prominent spike at $z\simeq 0.58$. 
Since the overdensity is compact in two dimensions on the sky and also in redshift this
is likely to be a cluster rather than a filament of large scale structure.
For the 16 galaxies we find  $z=0.574 \pm 0.008$ corresponding to a velocity dispersion $\sigma_v =  c \, \sigma_z/(1+z)$ $=$ 1500 km s$^{-1}$. This is typical of a rich cluster of galaxies and we find it is insensitive to the exact choice of radius. These parameters are consistent with the 
cluster containing the QSO absorber reported by T99 on it's outskirts.

 \section{Summary}
 
To summarize the paper:

\begin{enumerate}
\item We present a significant new set of photometric and spectroscopic
data on the HDF-S.
\item We have demonstrated a new mode of spectroscopic data taking
  with high-multiplex per unit area on the sky using the technique of
  `microslits' in conjunction with nod \& shuffle.
\item The spectroscopic catalog presented here is the third large
  catalog of redshifts in the HDF-S and flanking fields. 53 redshifts are obtained of
  which 45 are new.
\item Comparison with existing photometric redshifts based on WFPC-2
  data show these to be in reasonable agreement ($\sigma_z /(1+z)=0.03$) for $z<1$ galaxies,
  albeit with small number statistics.
\item We present strong evidence for a rich cluster, compact in all three spatial dimensions,
 at $z\simeq 0.58$ flanking
the WFPC2 pointing and which very likely contains the known absorbing galaxy
  of the QSO at this redshift.
\end{enumerate}

Preliminary versions of this catalog have already been used for scientific 
studies using the HDF-S data. Mann et al. (2002) used it for
a study of the relationship between far-infrared and other estimates
of galaxy star-formation rates. Vanzella et al. (2002) used this catalog together with
theirs to estimate cosmological star-formations rate history.
The reader is encouraged to use this catalog for further studies of the HDF-S
and its associated QSO.

\section*{} 

Based on data from the Anglo-Australian Observatory without whose
dedicated staff the LDSS upgrade project could not have proceeded. Nod
\& shuffle with LDSS was inspired by the pioneering ideas of Joss Bland-Hawthorn
to whom KG is indebted for many useful discussions. We
would like to especially thank Lew Waller, Tony Farrell and John
Barton for their technical work. We would also like to thank Roberto
Abraham for his help riding in the AAT Prime Focus cage during these
observations, Chris Tinney for observing support,
Pippa Goldschmidt for her help with the imaging
reductions and Sam Barden (NOAO), Jims Arns and Bill Coulburn
(Kaiser) for their help with the VPH grating design and manufacture.

\onecolumn

\begin{table}
\caption{Photometric Catalog (electronic edition only)}
\bigskip\tiny\tabcolsep=2pt
\begin{tabular}{llcccccccccccccc}
\hline\hline
 Phot ID & Coordinates & Area $^a$ &  Type Flag$^b$ & $R$$^c$ &  $Rerr$ &  R Class$^d$ &   R Mask$^e$&   R PHOT$^f$ & R Flag$^g$ &  $B$$^c$ &  $Berr$ &  B Class$^d$ &  B Mask$^e$ &   B PHOT$^f$ & B Flag$^g$\\
   & (J2000)     &           &                & (mag)   & (mag) & & & & & (mag)  &  (mag)   & & & & \\
   \hline
   1 &   22 32 28.10  $-$60 34 20.3 &  WF  &    G &  24.858 &   0.238 &   0.670 &    0& PH &  24&- & - & - &  -& -& -\\
    2 &   22 32 28.18  $-$60 34  9.7 &  WF  &    G &  25.210 &   0.264 &   0.668 &    0& PH &  16&- & - & - &  -& -& -\\
    3 &   22 32 28.25  $-$60 34 32.9 &  WF  &    G &  22.840 &   0.076 &   0.022 &    0& PH &  24&- & - & - &  -& -& -\\
    4 &   22 32 28.26  $-$60 33 52.0 &  WF  &    G &  23.884 &   0.126 &   0.025 &    0& PH &  24&- & - & - &  -& -& -\\
    5 &   22 32 28.26  $-$60 34 39.3 &  WF  &    G &  22.382 &   0.059 &   0.990 &    0& PH &  24&- & - & - &  -& -& -\\
    6 &   22 32 28.31  $-$60 34  0.9 &  WF  &    G &  24.186 &   0.171 &   0.355 &    0& PH &  24&- & - & - &  -& -& -\\
    7 &   22 32 28.36  $-$60 35  5.7 &  WF  &    G &  22.613 &   0.091 &   0.147 &    0& PH &  16&- & - & - &  -& -& -\\
    8 &   22 32 28.44  $-$60 33  9.6 &  WF  &    G &  22.608 &   0.048 &   0.983 &    0& PH &  26&- & - & - &  -& -& -\\
    9 &   22 32 28.48  $-$60 33 29.8 &  WF  &    G &  25.237 &   0.429 &   0.619 &    0& PH &  16&- & - & - &  -& -& -\\
   10 &   22 32 28.54  $-$60 34 15.2 &  WF  &    G &  23.737 &   0.191 &   0.685 &    0& PH &   2&- & - & - &  -& -& -\\
   11 &   22 32 28.57  $-$60 32 43.0 &  WF  &    G &  24.462 &   0.222 &   0.464 &    0& PH &  24&- & - & - &  -& -& -\\
   12 &   22 32 28.66  $-$60 34 49.7 &  WF  &    G &  23.862 &   0.221 &   0.477 &    0& PH &   0&- & - & - &  -& -& -\\
   13 &   22 32 28.67  $-$60 33 43.2 &  WF  &    G &  25.611 &   0.428 &   0.666 &    0& PH &   0&- & - & - &  -& -& -\\
   14 &   22 32 28.71  $-$60 33 57.5 &  WF  &    G &  24.058 &   0.214 &   0.657 &    0& PH &   2&- & - & - &  -& -& -\\
   15 &   22 32 28.72  $-$60 33 25.2 &  WF  &    G &  25.763 &   0.358 &   0.808 &    0& PH &   0&- & - & - &  -& -& -\\
   16 &   22 32 28.75  $-$60 32 11.9 &  WF  &    G &  23.927 &   0.180 &   0.782 &    0& PH &  24&- & - & - &  -& -& -\\
   17 &   22 32 28.78  $-$60 34  4.5 &  WF  &    G &  25.782 &   0.228 &   0.796 &    0& PH &   0&- & - & - &  -& -& -\\
   18 &   22 32 28.80  $-$60 32 37.9 &  WF  &    G &  23.950 &   0.210 &   0.065 &    0& PH &  16&- & - & - &  -& -& -\\
   19 &   22 32 28.81  $-$60 33 18.1 &  WF  &    G &  21.377 &   0.027 &   0.382 &    0& PH &  24&- & - & - &  -& -& -\\
   20 &   22 32 28.86  $-$60 33 59.4 &  WF  &    G &  24.262 &   0.222 &   0.095 &    0& PH &   3&- & - & - &  -& -& -\\
   21 &   22 32 28.94  $-$60 31 13.6 &  WF  &    G &  25.251 &   0.326 &   0.817 &    0& PH &  24&- & - & - &  -& -& -\\
   22 &   22 32 28.98  $-$60 34 34.2 &  WF  &    G &  24.851 &   0.206 &   0.930 &    0& PH &   0&- & - & - &  -& -& -\\
   23 &   22 32 29.01  $-$60 30 55.5 &  WF  &    G &  25.400 &   0.240 &   0.737 &    0& PH &  24&- & - & - &  -& -& -\\
   24 &   22 32 29.02  $-$60 33 22.4 &  WF  &    G &  23.591 &   0.154 &   0.025 &    0& PH &   0&- & - & - &  -& -& -\\
   25 &   22 32 29.03  $-$60 33 53.8 &  WF  &    G &  24.184 &   0.253 &   0.508 &    0& PH &   0&- & - & - &  -& -& -\\
   26 &   22 32 29.06  $-$60 34 14.2 &  WF  &    S &  23.486 &   0.082 &   0.977 &    0& PH &   0&- & - & - &  -& -& -\\
   27 &   22 32 29.08  $-$60 33 42.0 &  WF  &    G &  25.011 &   0.321 &   0.731 &    0& PH &   0&- & - & - &  -& -& -\\
   28 &   22 32 29.09  $-$60 30 46.7 &  WF  &    G &  24.048 &   0.174 &   0.339 &    0& PH &  24&- & - & - &  -& -& -\\
   29 &   22 32 29.14  $-$60 31  6.9 &  WF  &    G &  23.081 &   0.095 &   0.212 &    0& PH &  24&- & - & - &  -& -& -\\
   30 &   22 32 29.14  $-$60 32 48.6 &  WF  &    G &  20.725 &   0.016 &   0.034 &    0& PH &   0&- & - & - &  -& -& -\\
   31 &   22 32 29.23  $-$60 32 35.5 &  WF  &    G &  21.374 &   0.025 &   0.983 &    0& PH &   0&- & - & - &  -& -& -\\
   32 &   22 32 29.23  $-$60 33 33.3 &  WF  &    G &  23.614 &   0.116 &   0.506 &    0& PH &   0&- & - & - &  -& -& -\\
   33 &   22 32 29.23  $-$60 30 14.2 &  WF  &    G &  22.472 &   0.050 &   0.933 &    0& PH &  24&- & - & - &  -& -& -\\
   34 &   22 32 29.31  $-$60 34  0.3 &  WF  &    G &  24.950 &   0.319 &   0.534 &    0& PH &   0& 23.651 &   0.099 &   0.014 &    0& PH &  27\\
   35 &   22 32 29.32  $-$60 30 53.8 &  WF  &    G &  23.176 &   0.100 &   0.836 &    0& PH &  16&- & - & - &  -& -& -\\
   36 &   22 32 29.32  $-$60 31  4.0 &  WF  &    G &  24.095 &   0.204 &   0.631 &    0& PH &  18&- & - & - &  -& -& -\\
   37 &   22 32 29.32  $-$60 32 11.2 &  WF  &    S &  21.559 &   0.036 &   0.990 &    0& PH &   0&- & - & - &  -& -& -\\
   38 &   22 32 29.33  $-$60 34 37.1 &  WF  &    G &  24.646 &   0.187 &   0.725 &    1& PH &   0& 23.287 &   0.099 &   0.011 &    0& PH &  24\\
   39 &   22 32 29.34  $-$60 29 53.6 &  WF  &    G &  24.708 &   0.184 &   0.922 &    0& PH &  24&- & - & - &  -& -& -\\
   40 &   22 32 29.35  $-$60 34 10.3 &  WF  &    G &  24.263 &   0.150 &   0.738 &    0& PH &   0& 22.977 &   0.069 &   0.010 &    0& PH &  27\\
   41 &   22 32 29.35  $-$60 30 10.1 &  WF  &    G &  24.097 &   0.168 &   0.126 &    0& PH &  24&- & - & - &  -& -& -\\
   42 &   22 32 29.37  $-$60 29 49.9 &  WF  &    G &  25.378 &   0.254 &   0.767 &    0& PH &  24&- & - & - &  -& -& -\\
   43 &   22 32 29.38  $-$60 33 48.8 &  WF  &    G &  24.742 &   0.197 &   0.847 &    0& PH &   0& 22.404 &   0.055 &   0.000 &    0& PH &  27\\
   44 &   22 32 29.43  $-$60 29 30.7 &  WF  &    G &  23.845 &   0.132 &   0.973 &    0& PH &  24&- & - & - &  -& -& -\\
   45 &   22 32 29.46  $-$60 31 16.4 &  WF  &    G &  22.005 &   0.034 &   0.986 &    0& PH &   0&- & - & - &  -& -& -\\
   46 &   22 32 29.50  $-$60 32 43.3 &  WF  &    G &  22.474 &   0.046 &   0.758 &    0& PH &   0&- & - & - &  -& -& -\\
   47 &   22 32 29.53  $-$60 35  0.3 &  WF  &    S &  15.535 &   0.002 &   1.000 &    1& PH &  15& 16.560 &   0.002 &   1.000 &    0& PH &  30\\
   48 &   22 32 29.57  $-$60 32 29.9 &  WF  &    G &  23.547 &   0.126 &   0.029 &    0& PH &   0&- & - & - &  -& -& -\\
   49 &   22 32 29.57  $-$60 30 29.3 &  WF  &    G &  23.894 &   0.144 &   0.086 &    0& PH &   2&- & - & - &  -& -& -\\
   50 &   22 32 29.57  $-$60 30 39.6 &  WF  &    G &  24.757 &   0.203 &   0.870 &    0& PH &   0&- & - & - &  -& -& -\\
 \hline
\end{tabular}
\footnotesize
\bigskip

Notes:
Merged $B$ and $R$ selected catalog. Sample only. Full table
accessible from electronic edition and from the WWW at {\tt http://www.aao.gov.au/hdfs/Redshifts}. 
Dashed entries throughout this table denote missing data primarily due to 
the $B$ band data being less deep than the $R$ band but also slightly 
different depth and coverage at the edge of the fields and differences in 
extraction due to the presence of strong straylight features.
This table is not cleaned for sources lying in masked areas (i.e. in 
vignetted regions or along sites where stray light is strong) but the 
mask flag is included in the table.
\\

Notes to the column headings:\\

$^a$ Flag to denote parent image of the detection - WF: AAT-WFPC2 field, 
ST: AAT-STIS field, WS: Overlap region, WMS: Overlap region but masked in 
one of the fields. \\

$^b$ Star/Galaxy indicator as per the classification described in the text 
S: star G:non-stellar. \\

$^c$ SExtractors {\tt MAG\_AUTO} photometry, '-' not-detected.\\

$^d$ SExtractors star galaxy classifier - 0: galaxy, 1: star, '-' 
not-detected.\\

$^e$ Masked sources are flagged 0, unmasked sources are flagged 1, '-'
not-detected.\\

$^f$ Flag to indicate data taken under photometric conditions `PH' (WF-B, 
WF-R, ST-R data) or non-photometric `NP' where calibration using common 
sources detected in the photometric (WF-B) and non-photmetric (ST-B) lying 
in the overlap region has been used.\\

$^g$ SExtractor's output flag, generally $<$8 is reliable.\\

\label{tab:phot}
\end{table}

\newpage

\begin{deluxetable}{lccccccl}
\tablecolumns{8} \tabletypesize{\footnotesize} 
\tablecaption{Spectroscopic redshift catalog. \label{tab-spec}}
\tablehead{ 
  \colhead{Slit ID\tablenotemark{a}} & 
  \colhead{RA (J2000)} & 
  \colhead{Dec (J2000)} & 
  \colhead{Phot ID\tablenotemark{b}} & 
  \colhead{$R$ mag\tablenotemark{c}} & 
  \colhead{$z$} &  
  \colhead{$Q$} & 
  \colhead{Features\tablenotemark{d}, (HDF-S deep coverage) \hspace{20em}  }  \\} 
\startdata
225     & 22:32:38.909 & $-$60:34:54.095 &         472 &    22.54 & 9.9999 & 0 & Neg. spectrum. Edge effected.            \\
224     & 22:32:39.329 & $-$60:31:23.074 &         488 &    16.21 & 0.0000 & 4 & Star               \\
221     & 22:32:40.078 & $-$60:33:24.599 &         533 &    21.43 & 0.0000 & 4 & Mstar               \\
220     & 22:32:40.322 & $-$60:33:10.012 &         545 &    21.34 & 0.6213 & 2 & HK, 4000\AA              \\
216     & 22:32:41.357 & $-$60:30:26.276 &         592 &    19.27 & 0.4250 & 4 & OII, Balmer$-$, HK, H$\beta+$            \\
211     & 22:32:43.392 & $-$60:33:51.782 &         685 &    19.91 & 0.0918 & 1 & Broad line - best guess H$\alpha+$          \\
207     & 22:32:44.611 & $-$60:32:45.334 &         740 &    23.59 & 9.9999 & 0 & em line at 5862\AA, no cuum          \\
206     & 22:32:44.873 & $-$60:30:55.537 &         756 &    20.01 & 0.5141 & 4 & HK, 4000\AA,G, H$\gamma-$, H$\beta-$,NaD            \\
204     & 22:32:45.475 & $-$60:34:19.207 &         793 &    21.19 & 0.4606 & 3 & H$\beta+$,OIII(4959+5007)+               \\
197     & 22:32:47.393 & $-$60:32:00.179 &         899 &    21.29 & 0.0000 & 4 & Mstar   (WFPC2)            \\
196     & 22:32:47.592 & $-$60:33:36.122 &         912 &    19.94 & 0.5803 & 4 & OII,HK,H$\delta-$,G,H$\gamma-$ (WFPC2)               \\
185     & 22:32:50.419 & $-$60:34:01.063 &        1052 &    17.75 & 0.0000 & 4 & Mstar (WFPC2)              \\
173     & 22:32:53.952 & $-$60:31:17.915 &        1263 &    20.07 & 0.5838 & 2 & HK,OII,OIII               \\
170     & 22:32:54.763 & $-$60:31:13.876 &        1317 &    20.74 & 0.5111 & 3 & HK,4000               \\
169     & 22:32:55.022 & $-$60:34:28.999 &        1331 &    22.48 & 0.3562 & 3 & H$\beta+$,OIII(4959+5007)               \\
162     & 22:32:56.890 & $-$60:32:12.142 &        1432 &    22.93 & 0.7525 & 2 & HK, weak abs             \\
160     & 22:32:57.449 & $-$60:33:06.325 &        1462 &    20.57 & 0.5823 & 4 & OII,Balmer$-$,HK,G,H$\beta+$,OIII (WFPC2)              \\
157     & 22:32:58.236 & $-$60:33:51.962 &        1509 &    22.50 & 0.7063 & 2 & HK,G,H$\beta+$ (WFPC2)              \\
156*    & 22:32:58.706 & $-$60:33:23.818 &        1532 &    22.61 & 9.9999 & 0 & NoID (WFPC2)              \\
155     & 22:32:59.321 & $-$60:31:19.873 &        1561 &    13.60 & 0.0000 & 4 & Star               \\
154*    & 22:32:59.554 & $-$60:30:52.744 &        1578 &    23.70 & 0.2706 & 2 & OII,OIII               \\
153     & 22:32:59.820 & $-$60:31:01.690 &        1589 &    23.95 & 0.1916 & 3 & HK,G,H$\delta-$,H$\gamma-$               \\
152     & 22:33:00.082 & $-$60:33:19.336 &        1607 &    23.48 & 0.5402 & 2 & HK (WFPC2)              \\
145     & 22:33:02.376 & $-$60:33:46.850 &        1710 &    22.56 & 0.6959 & 1 & OII+ (strong), poss OIII+  (WFPC2)          \\
144     & 22:33:02.666 & $-$60:32:14.050 &        1725 &    18.81 & 0.0000 & 4 & Mstar               \\
143     & 22:33:02.974 & $-$60:32:31.589 &        1741 &    16.15 & 0.0000 & 4 & Mstar               \\
140     & 22:33:03.706 & $-$60:32:48.192 &        1781 &    20.66 & 0.2501 & 3 & H$\alpha+$,OIII+               \\
133     & 22:33:05.976 & $-$60:33:50.576 &        1891 &    17.20 & 0.1733 & 4 & H$\beta+$,OIII+,Mgb,5268,H$\alpha+$,NII+,SII+              \\
128     & 22:33:07.495 & $-$60:32:50.557 &        1997 &    20.17 & 0.5130 & 4 & HK,H$\delta-$,H$\beta+$               \\
127     & 22:33:08.095 & $-$60:33:22.000 &        2030 &    17.13 & 0.0000 & 4 & Star               \\
126     & 22:33:08.558 & $-$60:32:15.076 &        2052 &    23.59 & 0.5857 & 2 & OII+,OIII+ (both weak)             \\
123     & 22:33:09.444 & $-$60:33:44.244 &        2116 &    22.12 & 0.8090 & 1 & strong em line assumd OII. maybe weak abs?        \\
118     & 22:33:11.078 & $-$60:33:13.043 &        2229 &    21.83 & 0.5815 & 1 & strong em line assumed OII+, maybe wk abs        \\
117     & 22:33:11.462 & $-$60:32:33.515 &        2255 &    21.53 & 0.4411 & 2 & HK, H$\delta-$, G             \\
116     & 22:33:11.676 & $-$60:33:57.107 &        2273 &    23.17 & 0.4515 & 3 & HK, H$\delta-$,G              \\
115     & 22:33:11.866 & $-$60:30:52.632 &        2292 &    21.37 & 0.1633 & 2 & H$\alpha-$,Mgb,5268               \\
109\dag & 22:33:13.286 & $-$60:31:11.921 &        2380 &    21.56 & 0.4666 & 3 & HK, 4000\AA, also 2nd set of lines z=0.3080 (H$\beta+$,OIII+,H$\alpha+$)!       \\
108     & 22:33:13.574 & $-$60:34:06.319 &        2397 &    18.61 & 0.5430 & 3 & HK, G, H$\gamma-$             \\
107     & 22:33:14.083 & $-$60:33:55.037 &        2427 &    19.95 & 0.4842 & 2 & HK, G, H$\gamma-$,             \\
106     & 22:33:14.292 & $-$60:32:32.687 &        2774 &    23.39 & 9.9999 & 0 & Poss. em line at 7856\AA           \\
105     & 22:33:14.623 & $-$60:33:01.890 &        2461 &    22.56 & 0.4654 & 2 & HK,4000\AA               \\
104     & 22:33:14.866 & $-$60:31:06.470 &        2483 &    23.67 & 9.9999 & 0 & Contaminated by 103             \\
103     & 22:33:15.070 & $-$60:31:14.376 &        2495 &    20.25 & 0.3083 & 3 & OII+, H$\beta+$, H$\alpha+$, OIII+            \\
102     & 22:33:15.394 & $-$60:30:58.532 &        2521 &    23.76 & 0.4760 & 2 & H$\delta-$, G, H$\gamma-$ (H$\beta-$ on bad col.) (5861\AA abs is contam from 101.)   \\
101     & 22:33:15.674 & $-$60:32:24.868 &        2552 &    13.78 & 0.0000 & 4 & Mstar               \\
92      & 22:33:18.499 & $-$60:34:39.112 &        2730 &    18.65 & 0.0991 & 4 & H$\alpha+$,NII+,SII+               \\
91      & 22:33:18.782 & $-$60:30:30.373 &        2743 &    23.07 & 0.8839 & 3 & HK, MgII$-$, FeII$-$, Misc abs.           \\
90      & 22:33:19.003 & $-$60:32:27.820 &        2769 &    20.66 & 0.0000 & 4 & Mstar               \\
89      & 22:33:19.212 & $-$60:31:57.706 &        2782 &    22.80 & 0.3676 & 2 & HK, G, Mgb             \\
85      & 22:33:20.414 & $-$60:33:41.476 &        2856 &    20.61 & 0.1186 & 4 & H$\alpha+$,NII+,SII+, H$\beta+$, OIII+             \\
84      & 22:33:20.633 & $-$60:34:08.094 &        2863 &    23.01 & 1.2184 & 2 & HK, H$\delta-$, 4000\AA             \\
83      & 22:33:20.830 & $-$60:34:35.069 &        2877 &    16.93 & 0.0000 & 4 & Star               \\
82      & 22:33:21.161 & $-$60:31:42.964 &        2899 &    23.23 & 0.3256 & 2 & HK               \\
79      & 22:33:22.013 & $-$60:30:43.362 &        2962 &    21.00 & 0.2072 & 3 & HK,G,H$\beta-$,Mg               \\
74      & 22:33:24.018 & $-$60:33:10.605 &        3088 &    17.74 & 0.0000 & 3 & Star               \\
73      & 22:33:24.216 & $-$60:33:52.908 &        3102 &    16.15 & 0.0000 & 4 & Mstar               \\
67      & 22:33:25.867 & $-$60:31:19.315 &        3204 &    23.05 & 0.5800 & 3 & OII+,OIII(5007)+               \\
66      & 22:33:26.221 & $-$60:32:05.943 &        3229 &    15.76 & 0.0000 & 4 & Star               \\
65      & 22:33:26.480 & $-$60:33:54.955 &        3249 &    21.19 & 0.4802 & 2 & H,K,H$\beta-$               \\
64      & 22:33:26.718 & $-$60:33:57.125 &        3265 &    23.56 & 0.6428 & 3 & H,K,G               \\
59      & 22:33:28.046 & $-$60:33:38.037 &        3329 &    19.79 & 0.0000 & 3 & Star               \\
58      & 22:33:28.297 & $-$60:34:58.747 &        3338 &    22.98 & 0.3363 & 3 & OIII(5007+4959)+,H$\beta+$,H$\alpha+$?               \\
57      & 22:33:28.935 & $-$60:35:01.540 &        3365 &    21.36 & 0.0000 & 3 & Faint Mstar              \\
55      & 22:33:29.436 & $-$60:34:20.543 &        3395 &    23.71 & 1.3150 & 1 & Single line OII+             \\
54      & 22:33:30.012 & $-$60:34:01.470 &        3417 &    21.27 & 9.9999 & 0 & Dominated by bizarre negative spectrum - ghost?         \\
52      & 22:33:30.731 & $-$60:34:35.116 &        3451 &    23.47 & 0.3335 & 2 & Strong OIII 5007+, weak 4959+, H$\beta+$, maybe H$\alpha+$        \\
50      & 22:33:31.232 & $-$60:33:43.897 &        3476 &    19.31 & 0.0000 & 4 & M-star               \\
48      & 22:33:31.669 & $-$60:33:41.857 &        3501 &    23.87 & 9.9999 & 0 & Dominated by sky object = bright M star!!        \\
39      & 22:33:34.176 & $-$60:32:09.698 &        3616 &    22.28 & 0.2304 & 3 & OIII(5007+4959)+, H$\beta+$, H$\alpha+$             \\
31      & 22:33:37.404 & $-$60:34:03.207 &        3764 &    18.67 & 0.0000 & 4 & M-star               \\
27*     & 22:33:38.787 & $-$60:33:10.315 &        3825 &    23.25 & 0.5812 & 2 & OII+, wk abs lines (STIS)           \\
25      & 22:33:39.499 & $-$60:32:58.523 &        3864 &    23.15 & 1.2680 & 1 & Strong line at 8451\AA: assumed OII+          \\
20      & 22:33:41.350 & $-$60:32:56.122 &        3945 &    21.24 & 0.0000 & 2 & H$\alpha+$,NII+,SII+?               \\
13      & 22:33:43.962 & $-$60:31:23.017 &        4069 &    22.06 & 0.9534 & 1 & OII+, H$\beta+$?              \\
9       & 22:33:45.029 & $-$60:34:42.559 &        4117 &    23.68 & 9.9999 & 0 & Possible wk em line at 6979\AA          \\
8       & 22:33:45.231 & $-$60:35:18.301 &        4127 &    21.98 & 0.4049 & 2 & OIII+, OII+?, H$\beta+$, G            \\
5       & 22:33:46.169 & $-$60:34:03.448 &        4169 &    19.78 & 0.0000 & 4 & M star              \\
4*      & 22:33:46.439 & $-$60:34:06.634 &        4176 &    23.72 & 0.2645 & 2 & H$\alpha-$,Mgb               \\
3*      & 22:33:46.742 & $-$60:34:32.123 &        4196 &    22.60 & 0.2505 & 2 & H$\alpha-$,H$\beta-$,poss. Mgb and 5268            \\
2*      & 22:33:46.910 & $-$60:32:34.236 &        4207 &    23.33 & 0.8816 & 1 & Strong line OII+? at 7012\AA.           \\
1*      & 22:33:47.346 & $-$60:31:34.447 &        4221 &    21.91 & 0.2703 & 3 & Broad H$\beta+$,H$\delta+$, H$\theta+$, weak (but clear) OIII(5007+4959)+         \\
\enddata
\tablenotetext{\ }{Note: only objects with IDs ($Q>0$) are shown. The full table including unidentified objects is available on the web site.}
\tablenotetext{*}{Spectrum contaminated by scattered light.}
\tablenotetext{\dag}{Spectrum appears to contain two superimposed objects at different redshifts.}
\tablenotetext{a}{Identification number in slit mask.}
\tablenotetext{b}{Identification number in photometric catalog (Table \ref{tab:phot}).}
\tablenotetext{c}{$R$-band magnitude (Vega) in preliminary version of photometric catalog used for selection.}
\tablenotetext{d}{Principle features used to identify spectrum. $\pm$ denotes emission/absorption.}
\end{deluxetable}

\end{document}